\documentclass[11pt,a4paper]{article}

\usepackage[margin=1in]{geometry}
\usepackage{graphicx}
\usepackage{authblk}

\usepackage[utf8]{inputenc}
\usepackage{mathtools}
\usepackage{float}
\usepackage{amsfonts}
\usepackage{amssymb}
\usepackage{amsmath}
\usepackage{booktabs}
\usepackage[hidelinks]{hyperref}
\usepackage[numbers]{natbib}
\usepackage{xspace}
\usepackage[dvipsnames]{xcolor}
\usepackage{comment}
\usepackage{todonotes}
\newcommand{\rev}[1]{{\color{blue}{#1}}}
\renewcommand{\rev}[1]{#1}

\usepackage{amsthm}

\theoremstyle{plain}
\newtheorem{theorem}{Theorem}
\newtheorem{lemma}{Lemma}
\newtheorem{corollary}{Corollary}

\theoremstyle{definition}
\newtheorem{definition}{Definition}

\newtheorem*{proof*}{Proof.}

\usepackage{algorithm}
\usepackage{algpseudocode}
\algtext*{EndWhile} 
\algtext*{EndIf}
\algtext*{EndFor}

\usepackage{tikz}
\usetikzlibrary{positioning,chains,fit,shapes,calc}
\usetikzlibrary{patterns}
\definecolor{mygrey}{RGB}{235,235,235}

\newcommand{\ECP}{\textup{ECP}\xspace}
\newcommand{\KP}{\textup{KP}\xspace}
\newcommand{\PKP}{\textup{LDKP}\xspace}
\newcommand{\PKPL}{\textup{LDKP-}$\ell$\xspace}

\newcommand{\GAP}{\textup{GAP}\xspace}

\newcommand{\ECPs}{\textsc{Equal-Cardinality Partition Problem}\xspace}
\newcommand{\KPs}{\textsc{Binary Knapsack Problem}\xspace}
\newcommand{\PKPs}{\textsc{Location-Dependent Knapsack Problem}\xspace}
\newcommand{\PPs}{\textsc{Partition Problem}\xspace}
\newcommand{\GAPs}{\textsc{Generalized Assignment Problem}\xspace}
\newcommand{\KPDs}{\textsc{Knapsack Problem}\xspace}
\newcommand{\TPPs}{\textsc{3-Partition Problem}\xspace}

\newcommand{\skpi}{S_{1}\xspace}
\newcommand{\skpii}{S_{2}\xspace}
\newcommand{\spkp}{T\xspace}

\newcommand{\As}{A}
\newcommand{\Ae}{A_\delta}
\newcommand{\Sz}{A^{\hspace{-0.8pt}*}_{\delta}}

\newcommand{\re}{R_{\delta}}
\newcommand{\fe}{f_{\delta}}

\newcommand{\Oh}{\mathcal{O}}
\newcommand{\opt}{\textup{OPT}}
\newcommand{\e}{\ensuremath{\varepsilon}}

\newcommand{\calg}{\mathcal{P}}
\newcommand{\G}[1]{P_{#1}}
\newcommand{\Gmin}{\G{min}}
\newcommand{\Vmax}{V_{max}}

\renewcommand{\L}{C}

\newcommand{\gonly}{p}
\newcommand{\gionly}{\gonly_i}
\newcommand{\gi}[1]{\gionly{\left(#1\right)}}
\newcommand{\gl}[1]{\left(\L-#1\right)v_{i}}

\newcommand{\sonly}{w}
\newcommand{\s}[1]{\sonly_{#1}}

\usepackage{todonotes}

\newcommand{\keywords}[1]{\par\addvspace{0.5em}\noindent\textbf{Keywords:} #1\par}
\newenvironment{acknowledgements}{\section*{Acknowledgements}}{}

\begin{document}

\title{The Location-Dependent Knapsack Problem
\thanks{This work was supported by S\~ao Paulo Research Foundation (FAPESP) grants
\mbox{\#2015/11937-9},      
\mbox{\#2016/23552-7},      
\mbox{\#2020/13162-2},
\mbox{\#2022/05803-3}, and
\mbox{\#2025/20320-7},
and National Council for Scientific and Technological Development (CNPq) grants
\mbox{\#425340/2016-3},     
\mbox{\#312186/2020-7},    
\mbox{\#311039/2020-0},
\mbox{\#312345/2023-2}, and
\mbox{\#404779/2025-5}. 
}
}

\author[1]{Mauro R. C. da Silva}
\author[1]{Lehilton L. C. Pedrosa}
\author[1]{Rafael C. S. Schouery}
\affil[1]{Institute of Computing, University of Campinas, Campinas, Brazil\\
\texttt{maurorcsc@gmail.com},
\texttt{lehilton@ic.unicamp.br},
\texttt{rafael@ic.unicamp.br}}

\date{}

\maketitle

\begin{abstract}
  In this paper, we present the \PKPs~(\PKP), a variant of the \KPs~(\KP) in which the contribution of an item~$i$ to the objective value varies according to a function $\gi{h_i}$ that depends on the location $h_i \ge 0$ of item~$i$ in the knapsack.
  Each choice of functions $\gi{\cdot}$ corresponds to a distinct version of the problem, possibly with new properties compared to \KP.
  For the particular case in which~$\gi{h_i}$ is a simple decreasing linear function for each item~$i$, we note that \PKP is not a generalization of \KP.
  In this case, we prove that \PKP is NP-hard and present a Fully Polynomial-Time Approximation Scheme~(FPTAS).
  Moreover, we address a more general setting, assuming certain monotonicity and continuity conditions on functions $\gi{\cdot}$, for which we show that \PKP is strongly NP-hard and present a Polynomial-Time Approximation Scheme~(PTAS).

\end{abstract}

\section{Introduction}\label{sec:intro}

In the classical \KPs~(\KP), the task is to pack items into a container of a given capacity. Since it might not be possible to add all items to the container, one chooses a subset that can be packed and maximizes the total value.
Formally, an instance of \KP consists of a knapsack of capacity~$\L > 0$, a set of items~${I = \{1, 2, \dots, n\}}$ and, for each item~$i$, a value~${v_i > 0}$ and a \rev{weight}~${\s{i} > 0}$. The objective is to find a subset~${S \subseteq I}$ that maximizes~${\sum_{i \in S}{v_i}}$ and that does not exceed the capacity of the knapsack, i.e.,~${\sum_{i \in S}{\s{i}} \leq \L}$~\citep{kellerer2004introduction}. This is one of the most well-known problems in computer science, having many variants and generalizations~\citep{Cacchiani22a, Cacchiani22b}.

\rev{In several of these variants, however, the contribution of an item depends not only on which items are packed, but also on where they are placed.
A natural example arises in scheduling, where the profit (or cost) of a task depends on its completion time~\citep{gawiejnowicz2020review, gawiejnowicz2020models}.
In that setting, the capacity corresponds to a time horizon, and selecting and ordering jobs to maximize total profit yields a knapsack-like problem in which each job's contribution varies with its location in the schedule.
Another application appears in advertising layout: platforms such as Google and Mercado Livre (a large Latin American marketplace) display ads in a banner of fixed height, chosen from a large pool of candidate ads.
The expected revenue of an ad decreases with its distance from the top of the banner~\citep{yuan2020unbiased, zhu2010novel}, so both the selection and the placement of ads matter.
Because the banner height is limited while the number of candidates may be large, neither unrestricted packing nor enumeration of all feasible layouts is adequate.}
To model these tasks, we introduce the \PKPs~(\PKP), a variant of \KP in which the profit of an item varies with the location in which it is placed.


Formally, an instance of \PKP is composed of a knapsack of capacity~${\L > 0}$ and a set of items~$I = \{1, 2, \dots, n\}$, where each item~$i$ has value~${v_i > 0}$, \rev{weight}~${\s{i} > 0}$ and a function~$\gi{h_i}$, where~${h_i \ge 0}$ is the location at which the top of~$i$ is placed in the knapsack and corresponds to the total \rev{weight} of items preceding~$i$.
Both value and \rev{weight} are positive integers.
The profit of an item~$i$ whose top is placed in the knapsack at location $h_i$ is given by $\G{i} = \gi{h_i}v_i$.
The objective is to find a subset~${S \subseteq I}$ and the locations to pack~$S$ that maximize the total profit and do not exceed the capacity of the knapsack. See an example in Figure~\ref{fig:exemplo_pkp}, and notice that the top of the knapsack corresponds to location $0$. Also, note that when~$\gionly$ is a positive constant function for every item~$i$, we have the~\KP.

\definecolor{bluegray}{rgb}{0.4, 0.6, 0.8}
\definecolor{myyellow}{HTML}{FFF5CE}
\definecolor{mediumseagreen}{rgb}{0.24, 0.7, 0.44}

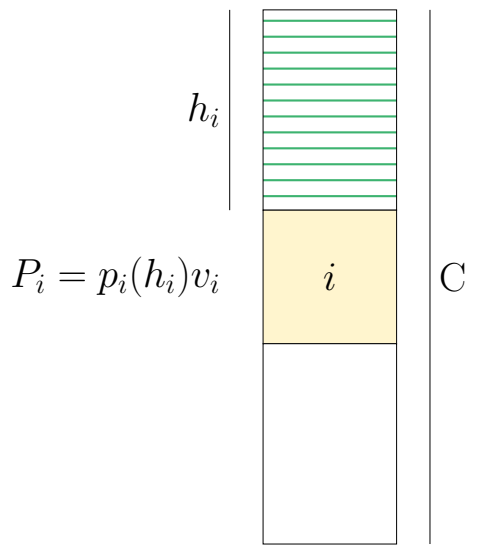
\begin{figure}[h]
  \centering
  \resizebox{.2\textwidth}{!}{%
    \begin{tikzpicture}

      \begin{scope}[start chain=going,node distance=10mm]
        \draw[black, pattern=horizontal lines, pattern color=mediumseagreen] (6, -4) rectangle ++(2, 3);
        \draw[black, fill=myyellow] (6, -6) rectangle ++(2, 2) node[pos=.5] {\LARGE$i$};
        \draw[black, fill=white] (6, -9) rectangle ++(2, 3);

        \draw (5.5, -5) node[left] {\LARGE$\G{i} = \gi{h_i}v_i$};


        \draw (8.5,-1) node[right, yshift=-4cm] {\LARGE\L} -- ++(0,-8);

        \draw (5.5,-1) node[left, yshift=-1.5cm] {\LARGE$h_i$} -- ++(0,-3);
      \end{scope}

    \end{tikzpicture}
  }
  \caption{Profit~$\G{i}$ of an item~$i$ when packed at location~$h_i$ of a knapsack.\label{fig:exemplo_pkp}}
\end{figure}

\rev{A closely related line of work is that of \citet{gawiejnowicz2023knapsack}, who consider knapsack problems in which an item's weight or profit depends on its index in the packing sequence.
These problems differ from~\PKP, where profits vary with the location in the knapsack rather than with the index in the sequence.
For classical~\KP and its many variants, we refer the reader to the recent surveys of~\citet{Cacchiani22a, Cacchiani22b}.}

\rev{In this work, we initiate the study of \PKP.
First, we consider a simple linear profit function, where $\gi{h_i} = \L-h_i$ for each item~$i$. We denote this particular case of~\PKP by~\PKPL. We show that the problem is NP-hard even for this case and present a \emph{Fully Polynomial-Time Approximation Scheme}~(FPTAS), that is, a family of algorithms that, for every~$\e > 0$, finds a solution within a factor of $(1 + \e)$ of the optimal solution in time polynomial in the instance size and in~$1/\e$.
Since a solution of \PKP includes the location in which items are packed, the objective function depends on the weights and values of packed items. This is distinct from \KP, whose objective function depends only on the values of packed items. To handle this difficulty, our FPTAS uses a recursive rounding approach.
Next, we present a \emph{Polynomial-Time Approximation Scheme}~(PTAS), a~$(1+\e)$-approximation whose running time is polynomial in the instance size for every fixed~$\e > 0$, considering more general functions, assuming certain monotonicity and continuity conditions, i.e., functions~${\gi{\cdot}}$ that are non-increasing and for all $0 < \delta < 1$, there is ${\e > 0}$ such that $\gi{h + \e} \geq (1 - \delta)\gi{h}$ for every~$h$. We also show that a PTAS is the best possible approximation algorithm for \PKP with such profit functions by proving that this problem is strongly NP-hard.}

\section{NP-hardness of \PKPL}

This section shows that~\PKPL is NP-hard.
When all items have the same value, \PKPL can be solved in polynomial time by greedily choosing the smallest items as they fit in the knapsack. Similarly, when all items have the same \rev{weight}, the problem can be solved in polynomial time by greedily choosing the items with the highest value while fitting in the knapsack.

Let $e_i = v_i/\s{i}$ be the \emph{efficiency} of an item~$i$. In what follows, we show that~\PKPL is NP-hard, even if all items have efficiency~$1$.

The \PPs is a classical NP-complete problem that consists of deciding whether it is possible to partition a set of positive integers into two parts with the same sum. The \ECPs~(\ECP) is a variant that requires that both parts of the solution have the same number of elements. Formally, given a set of indices~${I = \{1, 2, \dots, 2m\}}$ and a positive integer~$a_i$ for each~$i \in I$, we want to decide whether there is a subset~$S \subseteq I$ such that~${|S| = m}$ and~$\sum_{i \in S}{a_i} = \sum_{i \in I \setminus S}{a_i} = {\sum_{i \in I}{a_i}}/2$. As is the case of the \PPs, \ECP is also NP-complete~\cite{garey1979computers}.

Several variants of \KP are trivially NP-hard since they generalize or easily encode instances of the \PPs directly. Even more sophisticated problems can be demonstrated to be NP-hard as long as the objective function depends only on the total value of the packed items~\citep{kovalyov2010generic}. \rev{For \PKPL, however, standard reductions of this kind are not applicable, because maximizing the packed value (or the filled capacity) does not necessarily yield an optimal solution, as illustrated next.}

Both \KP and \PKPL have the same sets of inputs and feasible solutions.~Despite this similarity, the items of an optimal \KP solution do not necessarily form an optimal \PKPL solution for the same input even if all items have efficiency~$1$.

For example, consider the items described in Table~\ref{table:ce} and a knapsack of capacity~${\L = 10}$.~Optimal solutions for an instance of \KP with these items are ${\skpi = \{I_1, I_3, I_6\}}$ and~${\skpii = \{I_2, I_4, I_5\}}$. The values of these solutions when we consider the objective function of \PKPL are $5\cdot10 + 3\cdot5 + 2\cdot2 = 69$ and $4\cdot10 + 3\cdot6 + 3\cdot3 = 67$, respectively.
However, the optimal solution for \PKPL is~$\spkp = \{I_1, I_2\}$ and has profit ${5\cdot10 + 4\cdot5 = 70}$. Note that~$\spkp$ is not an optimal solution for \KP since~${\sum_{i \in \spkp}{v_i} = 9}$ and~${\sum_{i \in \skpi}{v_i} = \sum_{i \in \skpii}{v_i} = 10}$. Moreover, note that $\skpi$ and~$\skpii$ are also solutions for the \PPs in this instance, and~$\spkp$ is not a solution for that problem in the same instance. These solutions are shown in Figure~\ref{fig:exemplo}.

\begin{table}[h]
\centering
 {
  \def\arraystretch{0.8}\tabcolsep=10pt
  \begin{tabular}{l@{}l l l l l l l@{}l}
  \toprule
  & \textbf{$i$}  & \textbf{$1$} & \textbf{$2$} & \textbf{$3$} & \textbf{$4$} & \textbf{$5$} & \textbf{$6$} & \\
  \midrule
  & \textbf{$\s{i} = v_i$} & $5$  & $4$  & $3$  & $3$  & $3$  & $2$  & \\
  \bottomrule
  \end{tabular}
 }
\caption{Instance for the example of Figure~\ref{fig:exemplo}.\label{table:ce}}
\end{table}

\definecolor{babyblue}{rgb}{0.54, 0.81, 0.94}
\definecolor{babyblueeyes}{rgb}{0.63, 0.79, 0.95}
\definecolor{bluegray}{rgb}{0.4, 0.6, 0.8}
\definecolor{mediumseagreen}{rgb}{0.24, 0.7, 0.44}
\definecolor{darkcandyapplered}{rgb}{0.64, 0.0, 0.0}
\definecolor{deepcarrotorange}{HTML}{FFAE23}
\definecolor{mediumpurple}{HTML}{B363E6}
\definecolor{myyellow}{HTML}{FFF5CE}

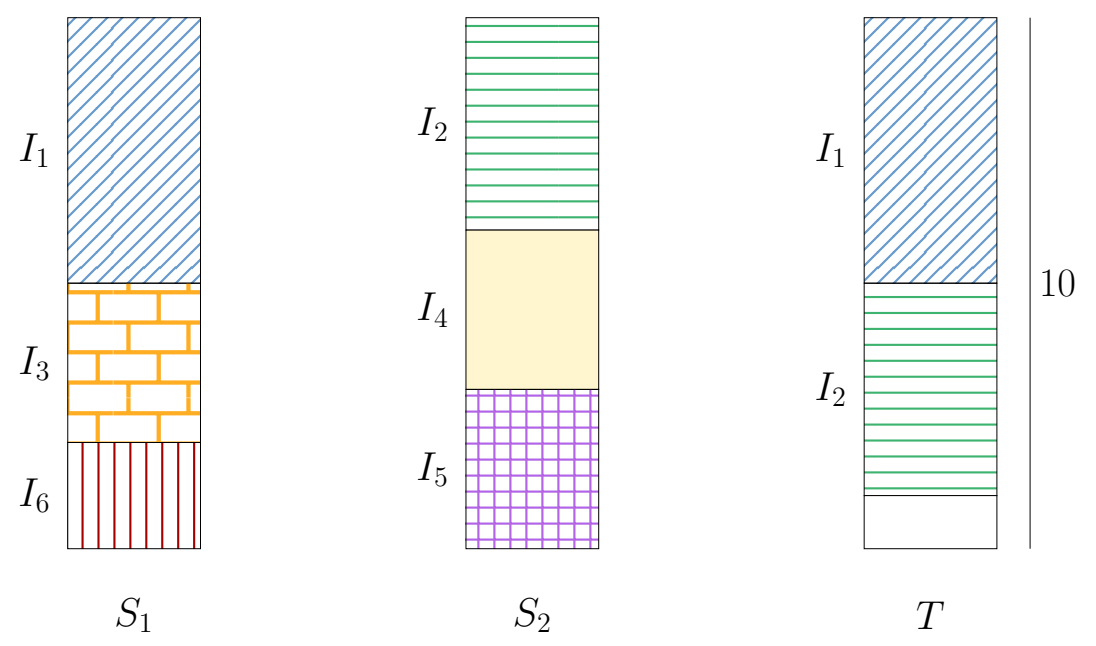
\begin{figure}[h]
  \centering
  \resizebox{.45\textwidth}{!}{%
    \begin{tikzpicture}

      \begin{scope}[start chain=going,node distance=10mm]
        \draw[black, pattern=north east lines, pattern color=bluegray] (-3, -5) rectangle ++(2, 4) node[pos=.5, xshift=-1.5cm] {\LARGE$I_1$};
        \draw[black, pattern=bricks, pattern color=deepcarrotorange] (-3, -7.4) rectangle ++(2, 2.4) node[pos=.5, xshift=-1.5cm] {\LARGE$I_3$};
        \draw[black, pattern=vertical lines, pattern color=darkcandyapplered] (-3, -9) rectangle ++(2, 1.6) node[pos=.5, xshift=-1.5cm] {\LARGE$I_6$};

        \draw[black, pattern=horizontal lines, pattern color=mediumseagreen] (3, -4.2) rectangle ++(2, 3.2) node[pos=.5, xshift=-1.5cm] {\LARGE$I_2$};
        \draw[black, fill=myyellow] (3, -6.6) rectangle ++(2, 2.4) node[pos=.5, xshift=-1.5cm] {\LARGE$I_4$};
        \draw[black, pattern=grid, pattern color=mediumpurple] (3, -9) rectangle ++(2, 2.4) node[pos=.5, xshift=-1.5cm] {\LARGE$I_5$};

        \draw[black, pattern=north east lines, pattern color=bluegray] (9, -5) rectangle ++(2, 4) node[pos=.5, xshift=-1.5cm] {\LARGE$I_1$};
        \draw[black, pattern=horizontal lines, pattern color=mediumseagreen] (9, -8.2) rectangle ++(2, 3.2) node[pos=.5, xshift=-1.5cm] {\LARGE$I_2$};
        \draw[black, fill=white] (9, -9) rectangle ++(2, 0.8);

        \draw (11.5,-1) node[right, yshift=-4cm] {\LARGE$10$} -- ++(0,-8);

        \draw (-2, -10) node[] {\LARGE$\skpi$};

        \draw (4, -10) node[] {\LARGE$\skpii$};

        \draw (10, -10) node[] {\LARGE$\spkp$};

      \end{scope}

    \end{tikzpicture}
  }
  \caption[Examples of solutions for \KP and \PKPL.]{Examples of solutions for \KP and \PKPL~consi\-dering the items in Table~\ref{table:ce}.\label{fig:exemplo}}
\end{figure}

Next, we show how to reduce \ECP to \PKPL{}\@. Given an instance of \ECP, let~${A = {\sum_{i \in I}{a_i}}/2}$. \rev{We may assume, without loss of generality, that~$a_i \le A$ for every~$i \in I$ as, otherwise, $a_i$ cannot be part of a solution to \ECP.} We create an instance of \PKPL as follows. Let~${M = 8A^2}$. Define the knapsack's capacity as~${\L = mM + A}$ and, for each $i \in I$, create an associated item whose \rev{weight} and value are equal to~$\s{i} = v_i = M + a_i$. The idea behind this reduction is to have very large items such that an integer in the original instance is much smaller than the \rev{weight} of the corresponding item. This makes all items very similar in \rev{weight}.

Let~${S \subseteq I}$ be a solution for \PKPL~considering a reduced instance and suppose, without loss of generality, that $S = \{1, 2, \dots, r\}$. Also, define $a(S) = \sum_{i \in S}a_i$ and~${a^2(S) = \sum_{i \in S }a_i^2}$.

\rev{Although the profit of each item depends on its location within the knapsack, every item in this construction satisfies~$\s{i} = v_i$. Exchanging two adjacent items~$i$ and~$j$ in any packing changes the total value by~${-\s{j}v_i + \s{i}v_j} = {-\s{i}\s{j} + \s{i}\s{j}} = 0$ (cf.\ the proof of Lemma~\ref{lem:pre1}), so the total value of a solution is unaffected by the packing order and is determined only by the set of packed items. This is why, in what follows, we write the value of a solution as a function~$f(S)$ of the set~$S$ alone.}

\rev{In particular, for such a solution~$S = \{1, \dots, r\}$,
\[
f(S)
=
\L \sum_{i \in S} \s{i}
- \frac{1}{2}\Bigl(\sum_{i \in S} \s{i}\Bigr)^{2}
+ \frac{1}{2}\sum_{i \in S} \s{i}^{2},
\]
since~$\sum_{i=1}^{r} \s{i} \sum_{j=1}^{i-1} \s{j} = \tfrac{1}{2}\bigl[(\sum_{i \in S} \s{i})^2 - \sum_{i \in S} \s{i}^2\bigr]$. Thus, $f(S)$ can be seen as a function that depends only on the values of~$\sum_{i \in S} \s{i}$ and~$\sum_{i \in S} \s{i}^2$. By substituting~$\s{i} = M + a_i$ for every~$i \in S$ into this identity, one obtains the closed form of Lemma~\ref{lem:statement0} below.
Note that the dominant terms of this closed form depend on the number of items~$r = |S|$.}

The proofs of the following lemmas are in Appendix~\ref{appendix:B}.

\begin{lemma}\label{lem:statement0}
 The value of a solution $S$ with $r$ items is
 \begin{footnotesize}
  \begin{flalign*}
 f(S) = \frac{1}{2}\left( r M \left( M - rM + 2mM + 2A\right) + 2 \left(\left(1 - r + m\right)M + A\right) a(S) - {a(S)}^2 + a^2(S)\right).
  \end{flalign*}
 \end{footnotesize}
\end{lemma}

\begin{lemma}\label{lem:3}
\rev{Every feasible solution~$S$ satisfies~$|S| \le m$. Moreover, every optimal solution has exactly~$m$ items.}
\end{lemma}

\begin{theorem}\label{thm:1}
\PKPL is NP-hard even when restricted to instances where all items have efficiency~$1$.
\end{theorem}

\begin{proof}
 \rev{In the decision version of \PKPL, given an instance of \PKPL and a parameter~$k$, one wants to decide whether a solution with value at least~$k$ exists.
 Let~$C_0 = mM(M + mM + 2A)$ and define
\[
 k = \frac{1}{2}\bigl(C_0 + 2(M + A)A - A^2\bigr).
\]
 We show that the instance of \ECP is ``yes'' if and only if the decision version of~\PKPL is ``yes''.

 First suppose that the instance of \ECP is ``yes'', and let~$S$ be a set of~$m$ indices with~$a(S) = A$. By Lemma~\ref{lem:statement0},
\[
 f(S)
 = \frac{1}{2}\bigl(C_0 + 2(M + A)A - A^2 + a^2(S)\bigr)
 \ge k,
\]
 since~$a^2(S) \ge 0$.

 Conversely, suppose that the decision version of \PKPL is ``yes'', and let~$S^*$ be an optimal solution with~$f(S^*) \ge k$. By Lemma~\ref{lem:3},~$|S^*| = m$. Feasibility implies~$a(S^*) \le A$. If~$a(S^*) < A$, then~$a(S^*) \le A - 1$ because the values~$a_i$ are positive integers. Using Lemma~\ref{lem:statement0} and~$a^2(S^*) \le {a(S^*)}^2$, we obtain
\[
 f(S^*)
 \le \frac{1}{2}\bigl(C_0 + 2(M + A)\,a(S^*)\bigr)
 \le \frac{1}{2}\bigl(C_0 + 2(M + A)(A - 1)\bigr).
\]
 Since~$M = 8A^2$,
\[
 k - \frac{1}{2}\bigl(C_0 + 2(M + A)(A - 1)\bigr)
 = M + A - \frac{1}{2}A^2
 > 0,
\]
 so~$f(S^*) < k$, a contradiction. Therefore~$a(S^*) = A$, and the instance of \ECP is ``yes''.}
\end{proof}

\section{Algorithms for \PKPL}\label{sec:fptas}

In this section, we present a pseudo-polynomial time algorithm and an FPTAS for~\PKPL{}\@.
Recall that the \emph{efficiency} of an item~$i$ is given by~${e_i = v_i/\s{i}}$.
\rev{In Lemma~\ref{lem:pre1}, we observe that a solution of \PKPL can be represented by the set of packed items alone, as we show that sorting the items in non-increasing order of efficiency maximizes the total profit.}
However, note that this does not mean that an optimal solution for \KP is also an optimal solution for \PKPL (as shown in the example of Figure~\ref{fig:exemplo}).

\begin{lemma}\label{lem:pre1}
 Let~$S$ be a solution for \PKPL that packs items~${I' \subseteq I}$ in an optimal order. Then the items are packed in non-increasing order of efficiency.
\end{lemma}

\begin{proof}
 \rev{Suppose, for a contradiction, that~$S$ packs $I'$ in an order that is not sorted in non-increasing order of efficiency. Thus, there are adjacent items~$i$ and~$j$ with~${h_i < h_j}$ and~${e_i < e_j}$.
 We have that ${v_i/\s{i} < v_j/\s{j}}$ and ${\s{j} v_i < \s{i} v_j}$.
 Note that, after exchanging the order of~$i$~and~$j$, the profit of items placed above~$i$ and below~$j$ remains unchanged. The profit of~$i$ changes from~$(\L-h_i)v_i$ to~$(\L-h_i - \s{j})v_i$ and the profit of~$j$ changes from~$(\L-h_j)v_j$ to $(\L - h_j + \s{i})v_j$. The difference between the new and previous profits is}
 \begin{small}
 \begin{align*}
 (\L-h_i - \s{j})v_i - (\L-h_i)v_i + (\L - h_j + \s{i})v_j - (\L-h_j)v_j
 = -\s{j}v_i + \s{i}v_j > 0.
 \end{align*}
 \end{small}%
\rev{Thus, exchanging the order of~$i$ and~$j$ improves the solution's value, which contradicts the assumption that~$S$ is optimal for~$I'$. Therefore, $S$ packs $I'$ in non-increasing order of efficiency.}
\end{proof}

\subsection{A Pseudo-polynomial Time Algorithm for \PKPL}\label{subsec:pp}

We present a pseudo-polynomial time algorithm for \PKPL, similar to the dynamic programming algorithm of~\citet{ibarra1975fast} for \KP{}\@.
Let~$\Vmax = \max_{i = 1}^{n}{v_i}$ be the largest value of an item.
We define the \emph{\rev{weight}} of a solution~$S \subseteq I$ as the sum~$\sum_{i \in S}{\s{i}}$.

We assume that the items are sorted in a non-increasing order of efficiency.
\rev{The algorithm creates a matrix~$\As$ with dimensions $n \times n\L\Vmax$ where each entry~$\As(i, q)$ of the matrix corresponds to the minimum weight of a solution considering only the first~$i$ items and with total profit exactly~$q$. If no such solution exists, then~${\As(i, q) = \infty}$.}
The following recurrence computes each entry of the matrix:
\rev{%
\begin{small}
\[
 \As(i, q) =
 \begin{cases}
  \infty,
  & \text{if } i = 1 \text{ and } \L v_1 \neq q, \\
  \s{1},
  & \text{if } i = 1 \text{ and } \L v_1 = q,  \\
  \min \bigl(\As(i-1, q),\, M(i, q)\bigr),
  & \text{otherwise},
 \end{cases}
\]
\end{small}
where
\begin{small}
\[
 M(i, q)
 =
 \min \bigl\{
 \As(i-1, k) + \s{i}
 :
 0 \le k < q
 \text{ and }
 k + \gl{\As(i-1, k)} = q
 \bigr\}.
\]
\end{small}}
\vspace{-0.5cm}

One can show that this recurrence computes an optimal solution for \mbox{\PKPL}. The algorithm's running time is~${\Oh(\L^2\Vmax^2 n^3)}$, which is pseudo-polynomial in the instance size.
Although this algorithm is not polynomial, we use the recurrence as the basis for the FPTAS presented in Subsection~\ref{subsec:fptas}.


\subsection{An FPTAS for \PKPL}\label{subsec:fptas}

\rev{We obtain an FPTAS for~\PKPL by combining the dynamic program of Subsection~\ref{subsec:pp} with geometric rounding of objective values.}
Consider a constant~$\e$ such that $0 < \e < 1$, and define $\delta = \e/n$.
Recall that Lemma~\ref{lem:pre1} states that items of an optimal solution for \PKPL are packed in a non-increasing order of efficiency. Thus, as before, we assume that the indices are sorted accordingly.
\rev{For each number~${x \ge 1}$, define~$\re(x)$ as ${(1 + \delta )}^q$ for some integer~$q$ such that ${{(1 + \delta )}^q \le x < {(1 + \delta )}^{q+1}}$.
We say that a value~$x$ is \emph{rounded down} to $\re(x)$.
Rounding objective values to powers of~$1 + \delta$ is a standard technique in the design of FPTAS's~\citep{woeginger2000when,chan2018approximation}.
See, e.g.,~\citet{boeckmann2023approximating} for a recent application to nonlinear knapsack problems.
We use a dynamic programming approach similar to that of Subsection~\ref{subsec:pp}, but with this geometric rounding, which reduces the matrix size and, consequently, the running time.}

\medskip

The algorithm creates a matrix $\Ae$ with dimensions~$n \times \left\lceil\log_{1 + \delta }{(n\L\Vmax)}\right\rceil$ such that each entry is computed using the following recurrence:
\rev{%
\begin{small}
\begin{equation}\label{eq:rec:fptas}
 \Ae (i, q) =
 \begin{cases}
  \infty,
  & \text{if } i = 1 \text{ and } \re(\L v_1) < {(1 + \delta )}^q,  \\
  \s{1},
  & \text{if } i = 1 \text{ and } \re(\L v_1) \ge {(1 + \delta )}^q, \\
  \min \bigl(\Ae(i-1, q),\, M_\e(i, q)\bigr),
  & \text{otherwise},
 \end{cases}
\end{equation}
\end{small}
where
\begin{small}
\begin{multline*}
 M_\e(i, q)
 =
 \min \Bigl\{
 \Ae(i-1, k) + \s{i}
 :
 0 \le k < q \\
 \qquad\text{and }
 \re\big(\gl{\Ae(i-1, k)} + {(1 + \delta )}^k\big) \geq {(1 + \delta )}^q
 \Bigr\}.
\end{multline*}
\end{small}}\vspace{-0.5cm}

The idea behind this recurrence is to use a dynamic programming algorithm similar to the one presented in Subsection~\ref{subsec:pp}, but approximating the value of a solution rounded down to a power of~$1 + \delta$.
\rev{Each entry~$\Ae(i, q)$ of the matrix corresponds to the weight of a solution considering only the first~$i$ items and with a total profit of at least~${(1 + \delta )}^q$.}
If the recurrence~\eqref{eq:rec:fptas} cannot find a solution, it sets~${\Ae(i, q) = \infty}$.



Let~$S$ be a solution with~$m$ items. We denote by~$S_i$ the~$i$-th item of~$S$ in non-increasing order of efficiency.
The total profit of~$S$ is defined as ${f(S) = {\sum_{i \in S}\G{i}}}$.
The \emph{rounded total profit} of~$S$ is recursively defined as follows:
\begin{footnotesize}
\[
 \fe(S) =
 \begin{cases}
  0,                    & \mbox{if } m = 0,  \\
  \re(\G{S_m} + \fe(S \setminus \{S_m\})), & \mbox{if } m \ge 1.
 \end{cases}
\]
\end{footnotesize}

The recursive function $\fe$ receives a set~$S$ with~$m$ items and sums its items' profits, rounding the accumulated profit with function $\re$ after adding each item's profit. This process is similar to the calculation of recurrence~\eqref{eq:rec:fptas}.

We define $\Sz(i, q)$ as the minimum \rev{weight} of a solution $S$ considering only the first~$i$ items and with rounded total profit ${\fe(S) \geq {(1 + \delta )}^q}$.
If no such solution exists, then~${\Sz(i, q) = \infty}$.
Lemma~\ref{lem:otimo} shows that recurrence~\eqref{eq:rec:fptas} computes $\Sz(i, q)$; it finds the best solution if the total profit is computed by~$\fe$.

\begin{lemma}\label{lem:otimo}
 For every $\delta$, $i$ and~$q$, the dynamic programming of recurrence~\eqref{eq:rec:fptas} computes $\Ae{(i, q)}$ such that $\Ae{(i, q) = \Sz(i, q)}$.
\end{lemma}

\begin{proof}
 We will show the lemma by induction on~$i$.
 If $i = 1$, a solution~$S$ can only include item~$1$, thus
 ${\fe(S) = \re(\L v_{1})}$, if $|S| = 1$, and
 ${\fe(S) = 0}$, otherwise.
 In this case,
 \begin{footnotesize}
 \[
  \Ae\left(1, q\right)=
  \begin{cases}
   \s{1},  & \text{if } \re(\L v_{1}) \ge {(1 + \delta )}^q, \\
   \infty, & \text{otherwise}.
  \end{cases}
 \]
 \end{footnotesize}
 Thus, indeed,~${\Ae(1, q) = \Sz(1, q)}$.

 Let $i\ge 2$ and assume ${\Ae(i-1, k) = \Sz(i-1, k)}$, for every~${k < q}$.
 Note that, if $\Ae(i, q) \neq \infty$, then we can build a solution $S$ considering only the first~$i$ items with $\fe(S) \geq {(1 + \delta )}^q$ by induction, and then,~${\Ae(i, q) \geq \Sz(i, q)}$.
 From now on, we prove that ${\Ae(i, q) \leq \Sz(i, q)}$.

 If ${\Sz(i, q) = \infty}$, then the result follows. Then, assume that ${\Sz(i, q) \neq \infty}$ and that there is a solution $S^*$ considering only the first $i$ items with ${\fe(S^*)} \geq {{(1 + \delta )}^q}$.

 First, suppose that item~$i$ is not in $S^*$. This implies that ${\Sz(i-1, q)} = {\Sz(i, q)}$, and then~${\Ae(i-1, q)} = {\Sz(i, q)}$ since ${\Ae(i-1, q)} = {\Sz(i-1, q)}$ by the induction hypothesis.
 But~${\Ae(i, q) \leq \Ae(i-1, q)}$, thus we have~${\Ae(i, q)} \leq {\Sz(i, q)}$.

 Suppose the item~$i$ is in $S^*$, and let~$\G{i}$ be the (not rounded) profit of $i$ concerning solution~$S^*$.
 Also, let $k$ be the integer such that $\fe(S^* \setminus \{i\}) = {(1 + \delta )}^k$.
 Note that~${k < q}$, as, otherwise, ${\fe(S^* \setminus \{i\}) \ge {(1 + \delta )}^q}$, and we could remove~$i$ from~$S^*$, maintaining the rounded total profit at least~${(1 + \delta )}^q$ while decreasing the \rev{weight}.

 Observe that~${S^* \setminus \{i\}}$ is a solution considering only the first~${i - 1}$ items with rounded total profit at least~${{(1 + \delta )}^{k}}$ that has minimum \rev{weight}.
 If not, then one could decrease the \rev{weight} of~$S^*$ by replacing~${S^* \setminus \{i\}}$ with the set of items corresponding to~$\Sz(i-1, k)$.
 It follows that~${\sum_{t \in S^* \setminus \{i\}} \s{t} = \Sz(i-1, k)}$.
 Since the items are in non-increasing order of efficiency, we have
 \[
  \textstyle
  \G{i}
 = \gl{\sum_{t \in S^* \setminus \{i\}} \s{t}}
 = \gl{\Sz(i-1, k)}.
 \]
 By the induction hypothesis,~${\Ae(i-1, k) = \Sz(i-1, k)}$, then
 \[
  \textstyle
  \G{i}
 = \gl{\Ae(i-1, k)}.
 \]
 Thus,
 \begin{align*}
  \fe(S^*)
  & = \re(\G{i} + \fe(S^* \setminus \{i\}))      \\
  & = \re(\gl{\Ae(i-1, k)} + {(1 + \delta )}^{k}) \\
  & \ge {(1 + \delta )}^q.
 \end{align*}

 We conclude that the entry $\Ae(i-1, k)$ is considered in the recurrence when computing~${\Ae(i, q)}$, and thus
 \begin{align*}
  \Ae(i, q) \le \Ae(i-1, k) + \s{i} = \Sz(i-1, k) + \s{i} = \Sz(i, q).\tag*{\qedhere}
 \end{align*}
\end{proof}

By Lemma~\ref{lem:otimo}, for each entry $\Ae(i,q)$ of the matrix, the recurrence~\eqref{eq:rec:fptas} can compute a corresponding solution~$S$ considering only the first $i$ items with a rounded total profit $\fe(S) \ge {(1 + \delta)}^q$ that minimizes the \rev{weight}.
Let~$S$ be the solution computed by the algorithm corresponding to the matrix $\Ae(n,q)$ entry such that $\Ae(n,q) \le \L$ and $q$ is maximum.
$S$ is the solution output by the algorithm.
Using the previous lemma, one can compare the value of this solution with the optimal value for~\PKPL{}\@.

\begin{lemma}\label{lem:1fptas}
 Let~$S^*$ be an optimal solution for \PKPL and $S$ be the solution output by recurrence~\eqref{eq:rec:fptas}. Then,~${\fe(S) \ge \exp(-2\e) f(S^*)}$.
\end{lemma}

\begin{proof}
 Recall that the rounded value of a number~$x$ is $\re(x)$ and that $\re(x) \le x < (1+\delta) \re(x)$, thus we have~${\re(x) > (1 - \delta) x}$.

 Without loss of generality, assume $S^* = \{1, 2, \dots, m\}$.
 We show by induction on~$m$ that $\fe(S^*) \ge \sum_{i=1}^m {(1-\delta)}^i\G{m-i+1}$.

 For $m = 1$, we have $\fe(S^*) = \re(\G{1} + \fe(S^* \setminus \{1\})) = \re(\G{1}) \ge (1-\delta)\G{1}$ and the claim holds. Thus, let $m \ge 2$ and assume the claim holds for $m-1$.
 We have
 \begin{align*}
  \fe(S^*)
  & = \re(\G{m} + \fe(S^* \setminus \{m\}))                  \\
  & \ge (1-\delta) (\G{m} + \fe(S^* \setminus \{m\}))             \\
  & \ge (1-\delta) \left(\G{m} + \sum_{i=1}^{m-1} {(1-\delta)}^i\G{(m-1)-i+1}\right)   \\
  & = (1-\delta) \G{m} + (1-\delta)\sum_{i=2}^{m} {(1-\delta)}^{i-1}\G{m-i+1} \\
  & = \sum_{i=1}^m {(1-\delta)}^i\G{m-i+1}.
 \end{align*}

 Now, since $n \ge m$,
 \[
  \fe(S^*) \ge \sum_{i=1}^m {(1-\delta)}^i\G{m-i+1} \\
  \ge {(1-\delta)}^n\sum_{i=1}^m \G{m-i+1} \\
 = {(1-\delta)}^n f(S^*).
 \]

 Note that~${\fe(S) \geq \fe(S^*)}$, because~$S$ is optimal with respect to the rounded total profit~$\fe$. Thus, $\fe(S) \geq \fe(S^*) \geq {(1-\delta)}^n f(S^*)$.

 Substituting $\delta$ by $\frac{\e}{n}$, we get $\fe(S) \geq {\left(1-\frac{\e}{n}\right)}^n f(S^*)$.
 Since $n \ge 2$ and $\e < 1$, we have $\frac{\e}{n} \in [0, 1/2]$, and thus $\ln\left(1-\frac{\e}{n}\right) \ge -2\frac{\e}{n}$, using that $\ln(1-x) \ge -2x$ for every~${x \in [0, 1/2]}$.
 Multiplying both sides by~$n$ and exponentiating, we get
 \begin{align*}
  \fe(S) \geq {\left(1-\frac{\e}{n}\right)}^n f(S^*) \ge
  \exp(-2\e) f(S^*).\tag*{\qedhere}
 \end{align*}
\end{proof}

In Lemma~\ref{lem:poly}, we show that recurrence~\eqref{eq:rec:fptas} runs in polynomial time.
Combining Lemmas~\ref{lem:1fptas}~and~\ref{lem:poly}, we show that the dynamic programming~\eqref{eq:rec:fptas} is an FPTAS for~\PKPL{}\@.

\begin{lemma}\label{lem:poly}
 \rev{Recurrence~\eqref{eq:rec:fptas} can be computed in time~$\Oh\!\left(\frac{n^3\log^2{(n\L\Vmax)}}{\e^2}\right)$.}
\end{lemma}

\begin{proof}
\rev{Sorting the items by efficiency takes time $\Oh(n\log{n})$, which is polynomial. The matrix size is $\Oh(n \log_{ 1 + \delta }{(n\L\Vmax)})$ and each entry can be computed in time ${\Oh(\log_{1 + \delta }{(n\L\Vmax)})}$. Thus, as~$\delta = \e/n$, the algorithm complexity is}
\begin{align*}
 \Oh\left(n\log{n} + n \log_{1 + \frac{\e}{n}}^2{(n\L\Vmax)}\right)
 & =
 \Oh\left(\frac{n\log^2{(n\L\Vmax)}}{\log^2{\left(1 + \frac{\e}{n}\right)}}\right) \\
 & =
 \Oh\left(\frac{n\log^2{(n\L\Vmax)}}{{\left(\frac{\e}{2n}\right)}^2}\right)  \\
 & =
 \Oh\left(\frac{n^3\log^2{(n\L\Vmax)}}{\e^2}\right).
\end{align*}
\rev{The second equality holds since~${\e/n \in (0, 1)}$ and ${\log{(1 + x)} \geq x/2}$ for ${x \in [0, 1]}$.}%
\end{proof}

\begin{theorem}\label{thm:fptas}
 \rev{For every~$0 < \e' < 1/2$, the algorithm produces a solution of value at least~${(1-\e')\opt}$, where~$\opt$ is the value of an optimal solution, in time
 \[
 \Oh\!\left(\frac{n^3\log^2{(n\L\Vmax)}}{{\e'}^2}\right).
 \]
 In particular, there exists an FPTAS for~\PKPL{}\@.}
\end{theorem}

\begin{proof}
 \rev{Fix~$0 < \e' < 1/2$ and set~${\e = \frac{1}{2}\ln \left( 1/(1-\e') \right)}$, so that~${0 < \e < 1}$; run the algorithm of recurrence~\eqref{eq:rec:fptas} with this~$\e$.}
 By Lemma~\ref{lem:1fptas}, the output solution $S$ has value at least~${f(S) \ge \fe(S)} \ge \exp(-2\e) \opt$.
 \rev{Since~$\e = \frac{1}{2}\ln \left( 1/(1-\e') \right)$,}
 \begin{align*}
 f(S) \ge \exp(-2\e) \opt = (1 - \e') \opt.
 \end{align*}

 \rev{Moreover,}
 \begin{align*}
  \frac{1}{\e}
 = \frac{1}{\frac{1}{2}\ln \left( \frac{1}{1-\e'}\right)}
 = \frac{-2}{\ln (1-\e')}
  \le \frac{2}{\e'}.
 \end{align*}
 \rev{By Lemma~\ref{lem:poly}, the algorithm runs in time
 \[
 \Oh\!\left(\frac{n^3\log^2{(n\L\Vmax)}}{\e^2}\right)
 =
 \Oh\!\left(\frac{n^3\log^2{(n\L\Vmax)}}{{\e'}^2}\right),
 \]
 since~${1/\e \le 2/\e'}$. In particular, the running time is polynomial in the size of the instance and in~$1/\e'$, establishing that the algorithm is an FPTAS for~\mbox{\PKPL{}}\@.}
\end{proof}

\section{A More General Version of \PKP}\label{sec:ptas}

This section considers variants of \PKP with more general profit functions. In Subsection~\ref{subsec:def}, we describe the assumptions on the profit functions and give some examples. In Subsection~\ref{subsec:np-hardness-tardiness}, we show that \PKP is strongly NP-hard even for simple piecewise linear profit functions, and, in Subsection~\ref{subsec:ptas2}, we present a PTAS for relatively continuous profit functions.

Since, in the general \PKP, every item has a different profit function, we can define profit functions \rev{$\gionly$} and \rev{$\gonly_j$} for two items $i$ and $j$ with $e_i > e_j$ such that $v_j \gonly_j(h) > v_i \gi{h}$ for all~$h$. Then, an optimal solution for those two items will pack~$j$ before~$i$, contradicting Lemma~\ref{lem:pre1}, and we cannot use the FPTAS for this version of \PKP.

\subsection{Assumptions on the Profit Functions}\label{subsec:def}

We say that a function~$\gonly$ is \emph{continuous} if, for all $x_0$ and for all $\delta > 0$, there is $\e > 0$ such that ${|\gonly(x) - \gonly(x_0)| < \delta}$, for all $x$ such that ${|x - x_0| < \e}$.
A function~$\gonly$ is said to be \emph{uniformly continuous} if, for all $\delta > 0$, there is $\e > 0$ such that ${|\gonly(x) - \gonly(y)| < \delta}$, for all~$x$~and~$y$ such that~${|x - y| < \e}$.
In this work, we assume that the profit function for each item satisfies some notion of continuity, as defined next.

\begin{definition}\label{def:relcont}
 A non-increasing function~${\gonly \colon [0, \L] \rightarrow \mathbb{R}_{\ge 0}}$ is \emph{relatively continuous} if, for all~${0 < \delta < 1}$, there is~${0 < \e < 1}$ such that ${\gonly(h + \e) \geq (1 - \delta)\gonly(h)}$ for all~$h \in [0, \L]$.\footnote{\rev{For convenience, we set~${\gonly(h) = \gonly(\L)}$ for every~${h > \L}$, so that~$\gonly(h + \e)$ is well defined for all~${h \in [0, \L]}$ and~${\e > 0}$. This extension preserves both monotonicity and uniform continuity.}}
\end{definition}

\rev{Examples of functions with this property are non-increasing uniformly continuous functions with minimum value one, as stated in Theorem~\ref{thm:fun1}.}

\begin{theorem}\label{thm:fun1}
 If a function $\gonly \colon [0, \L] \rightarrow \mathbb{R}_{\ge 1}$ is non-increasing and uniformly continuous, then $\gonly$ is relatively continuous.
\end{theorem}

\begin{proof}
 Let~$0 < \delta < 1$.
 Since~$\gonly$ is uniformly continuous, there is~$\e' > 0$ such that ${|\gonly(x) - \gonly(y)| < \delta}$ for all~$x$ and~$y$ with~${|x - y| < \e'}$.
 Take any~$\e$ such that~${0 < \e < \min\{\e', 1\}}$.
 Since~$\gonly$ is non-increasing, ${0 \le \gonly(h) - \gonly(h + \e) < \delta}$ for all~$h$, and thus
 \[
  \gonly(h + \e) > \gonly(h) - \delta
  \geq
  \gonly(h) - \delta \gonly(h)
 = (1-\delta) \gonly(h).
 \]
 The second inequality holds because~$\gonly(h) \ge 1$ for all~$h \in [0, \L]$.
 \end{proof}

Linear functions are among the simplest examples of uniformly continuous functions. Other examples are the Lipschitz continuous functions and the $\alpha$-Hölder continuous functions. Also, any continuous function on the interval~$[0, \L]$ is also uniformly continuous \citep{spivak2019calculus}, thus, ${p_{\mathrm{quad}}(h)} = {(\L^2-h^2) + 1}$ and ${p_{\mathrm{sqrt}}(h)} = {(\sqrt{\L}-\sqrt{h}) + 1}$ are examples of non-increasing and uniformly continuous functions for which Theorem~\ref{thm:fun1} holds.

\bigskip

\rev{We say that a set~$\calg$ of relatively continuous functions is relatively continuous if, for every~$0 < \delta < 1$, there exists~$0 < \e < 1$ that satisfies the definition simultaneously for every function in~$\calg$.
Any finite set of relatively continuous functions, or any finite union of relatively continuous sets, has this property.
For each fixed set~$\calg$, we consider the variant of~\PKP in which each item~$i \in I$ is associated with some~$\gionly \in \calg$, and the profit of~$i$ at location~$h_i$ is~$\G{i} = \gi{h_i}v_i$.
If~$\calg$ is finite, an instance may simply store, for each item, the index of its function.
Finally, we assume each~$\gionly$ is continuous: although an optimal packing without wasted space uses only finitely many positions, the PTAS may introduce gaps that are multiples of~$\e$, and computing a nearest position consistent with the items' weights would otherwise require solving a~\KPDs.}

\subsection{Strong NP-hardness of \PKP}\label{subsec:np-hardness-tardiness}

\rev{In this subsection, we give a proof that~\PKP is strongly NP-hard, based on a reduction from the single-machine total weighted tardiness problem~$1||\sum \bar{w}_j T_j$.
The argument already applies to piecewise-linear profit functions with only two pieces.}

\rev{An instance of~$1||\sum \bar{w}_j T_j$ consists of~$n$ jobs, where each job~$j$ has a processing time~$\bar{p}_j > 0$, a due date~$d_j \ge 0$ and a weight~$\bar{w}_j > 0$.
A schedule is a permutation of the jobs on a single machine starting at time~$0$ with no idle time.
If job~$j$ completes at time~$\bar{C}_j$, its tardiness is~$T_j := \max\{0, \bar{C}_j - d_j\}$.
The goal is to minimize~$\sum_{j=1}^{n} \bar{w}_j T_j$.
This problem is strongly NP-hard~\citep{lenstra1977complexity}.}

\rev{%
Given such input data, we construct a corresponding instance of~\PKP as follows.
Set the knapsack capacity to~$\L = \sum_{j=1}^{n} \bar{p}_j$, so that packing every item fills the knapsack exactly.
We may assume that~$d_j \le \L$ for every job~$j$, since no job completes after time~$\L$; thus, replacing a due date larger than~$\L$ with~$\L$ does not increase tardiness.
For each job~$j$, create an item with weight~$\s{j} = \bar{p}_j$, value~$v_j = \bar{w}_j$, and profit function
\begin{equation*}
\gonly_j(h_j)
=
\min\bigl\{\L - d_j,\; \L - h_j - \bar{p}_j\bigr\} + \bar{p}_j + \L
=
\min\bigl\{2\L - d_j + \bar{p}_j,\; 2\L - h_j\bigr\}
\end{equation*}
for all~$h_j \in [0, \L]$.
Thus~$\gonly_j(h_j) = 2\L - d_j + \bar{p}_j$ whenever~$h_j \le d_j - \bar{p}_j$ (the job would complete on time), and~$\gonly_j(h_j) = 2\L - h_j$ whenever~$h_j > d_j - \bar{p}_j$ (the profit decreases linearly with the tardiness).
Each function~$\gonly_j$ is non-increasing and piecewise linear with two pieces, hence uniformly continuous, and it satisfies~${\L = \gonly_j(\L) \le \gonly_j(h_j) \le 2\L}$ for every~$h_j \in [0, \L]$.
Since~${\L \ge 1}$, Theorem~\ref{thm:fun1} implies that each~$\gonly_j$ is relatively continuous.
The additive term~${\bar{p}_j + \L}$ shifts the profit of item~$j$ by a fixed amount, and hence, as shown next, it changes the objective value only by a constant.}

\begin{theorem}\label{thm:np-hardness-tardiness}
\rev{\PKP is strongly NP-hard, even when every profit function is relatively continuous and piecewise linear with one constant piece and one decreasing linear piece.}
\end{theorem}

\begin{proof}
\rev{%
In the constructed instance, every optimal solution packs all items.
Indeed, since the profit functions are non-increasing, we may assume that a solution leaves no idle space, and, if some item~$j$ is not packed, we may place it at location~$h_j = \sum_{i \in S} \s{i}$, where~$S$ is the set of packed items.
This is feasible because~$h_j + \s{j} \le \sum_{i \in I} \s{i} = \L$, it does not change the location of any other item, and it increases the objective value by~$v_j \gonly_j(h_j) \ge v_j > 0$.

Packing all items without idle space is equivalent to choosing a permutation of the jobs: if item~$j$ starts at location~$h_j$, then its completion time in the corresponding schedule is~$\bar{C}_j = h_j + \bar{p}_j$.
For every item~$j$ we have
\begin{align*}
\gonly_j(h_j)
&=
\min\bigl\{\L - d_j,\; \L - h_j - \bar{p}_j\bigr\} + \bar{p}_j + \L \\
&=
\L - d_j - \max\bigl\{0,\, h_j + \bar{p}_j - d_j\bigr\} + \bar{p}_j + \L \\
&=
\L - d_j - \max\bigl\{0,\, \bar{C}_j - d_j\bigr\} + \bar{p}_j + \L =
2\L - d_j + \bar{p}_j - T_j.
\end{align*}
Therefore, the objective value of the~\PKP packing equals
\begin{align*}
\sum_{j=1}^{n} v_j \, \gonly_j(h_j)
&=
\sum_{j=1}^{n} \bar{w}_j \bigl(2\L - d_j + \bar{p}_j - T_j\bigr) \\
&=
\sum_{j=1}^{n} \bar{w}_j (2\L - d_j + \bar{p}_j)
-
\sum_{j=1}^{n} \bar{w}_j T_j.
\end{align*}
The first sum on the right-hand side is constant.
Hence, maximizing the~\PKP objective is equivalent to minimizing the total weighted tardiness~$\sum_{j} \bar{w}_j T_j$.

The construction is computed in polynomial time, and every number of the constructed instance is bounded by a polynomial in the numbers of the scheduling instance, so the reduction preserves strong NP-hardness.
Since~$1||\sum \bar{w}_j T_j$ is strongly NP-hard, the claim follows.
In particular, while~\PKPL is only weakly NP-hard, already the passage from a purely linear profit function to a piecewise linear function with two pieces makes~\PKP strongly NP-hard.}
\end{proof}

\rev{We remark that the constructed functions form a relatively continuous set: after scaling the capacity to one and dividing the profits by~$\L$, each of them is $1$-Lipschitz and takes values in~$[1, 2]$, and thus~${\e = \delta}$ satisfies Definition~\ref{def:relcont} simultaneously for all of them, over every scheduling instance.
Hence \PKP is strongly NP-hard already for a fixed relatively continuous set~$\calg$.}

\subsection{A PTAS for \PKP with Relatively Continuous Profit Functions $\calg$}\label{subsec:ptas2}

\rev{In the following, consider a fixed set of functions~$\calg$ that is relatively continuous as the set of profit functions, and consider an instance of the corresponding variant of \PKP.
Without loss of generality, we suppose that the knapsack's capacity is~$\L=1$.}
Our algorithm can be summarized in the following outline:
\begin{itemize}
\item For any $0 < \delta < 1$, define $\e \leq \delta$ such that $1/\e$ is an integer and $\gi{h + \e} \geq (1-\delta)\gi{h}$ for all profit functions $\gionly \in \calg$ and $h \in [0,1]$;
\item \rev{Enumerate every candidate set~$V \subseteq I$ with~$|V| \le 2/\e^2$ together with every ordering of its items;}
\item \rev{For each such candidate~$V$ and ordering, define intervals between consecutive items of~$V$ and approximate each interval length by a multiple of~$\e$;}
\item Fill each interval with items not in~$V$ using a PTAS for the \GAPs~(\GAP).
\end{itemize}

Let $\delta > 0$ be some fixed constant. In this subsection, the goal is to find a solution within a factor of~$1-\Oh(\delta)$ of the optimal value.
Since $\calg$ is relatively continuous, there exists a constant~$\e > 0$ with~$\e \le \delta$ such that $1/\e$ is an integer and, for every item $i \in I$, we have~${\gi{h + \e} \geq (1 - \delta)\gi{h}}$ for all $h \in [0,1]$.

In what follows, we denote by $(S^*, h^*)$ some fixed optimal solution, so each selected item~${i \in S^*}$ has start location $h^*_i$.
Assume that $|S^*| \ge 1/\e^2$, as otherwise, we could solve the instance exactly in polynomial time.

\rev{For the analysis, let~$V^*$ be the subset of items of~$S^*$ containing the~$1/\e^2$ items with the largest profits and every item with weight at least $\e^2$.
Note that, since a solution can have at most $1/\e^2$ items with weight at least $\e^2$, the cardinality of~$V^*$ is at most $2/\e^2$, which is bounded by a constant.
The algorithm does not know~$V^*$; however, since it enumerates all sets of size at most~$2/\e^2$ together with all orderings of their items, one of these candidates coincides with~$V^*$ and with the packing order of these items in~$(S^*, h^*)$.
In what follows, we analyze this successful iteration.}
For the sake of simplicity, assume that $V^* = \{1, 2, \dots, |V^*|\}$ and that the items are positioned in this order, as we could change the indices if this was not the case.

Since the profit functions are non-increasing, the space between consecutive items of~$V^*$ in the optimal solution contains items whose \rev{weight} is smaller than $\e^2$.
We say that each such space corresponds to an \emph{interval} of the knapsack.
The interval immediately before an item~${i \in V^*}$ is identified by index $j = i$, and the last interval is identified by index $j = |V^*| + 1$.
Also, let~${h_{|V^*|+1}^* = 1}$ and denote by $c_j^*$ the length of the interval~$j$, thus this interval ends at location~$h_j^*$ and corresponds to the set of locations~$[h_j^* - c_j^*, h_j^*)$.

Since we analyze the iteration corresponding to~$V^*$, the end location $h_j^*$ of each interval $j$ could be determined if we knew the vector $c^*$ with interval lengths.
We cannot enumerate all candidate vectors because the number of possibilities is exponential. Rather, for each interval $j$, we would like to find a number $c_j$ that is a multiple of $\e$, such that $c_j$ approximates~$c_j^*$ with an absolute difference of at most $\e$.
Because the knapsack has unit capacity, the number of candidate vectors for $c$ is bounded by a constant.

We define a vector of lengths~$c$ and a corresponding vector of end locations~$h$ by modifying the locations of items in $V^*$ iteratively.
First, define $c_1$ as the smallest multiple of $\e$ that is at least~$c^*_1$, and set~$h_1 = c_1$.
Then, for each interval $j \ge 2$ in order, define~$c_j$ as the smallest nonnegative multiple of $\e$ that is at least $h^*_j - (h_{j-1} + \s{j-1})$ and set $h_j = h_{j-1} + \s{j-1} + c_j$.
Intuitively, one may think of this procedure as ``pushing down'' the items of $V^*$.
The following lemma (proved in Appendix~\ref{appendix:B}) notes that the end locations do not change by much.

\begin{lemma}\label{lem:altura0}
 For every interval~$j$, we have ${0 \le h_j - h^*_j < \e}$.
\end{lemma}

The modified intervals have the key property that the cumulative length is at least as large as that of the optimal solution. This is formalized in the following corollary (proved in Appendix~\ref{appendix:B}).

\begin{corollary}\label{cor:cumulative0}
 For every interval~$j$, we have
${ \sum_{\ell = 1}^j c_\ell \ge \sum_{\ell = 1}^j c^*_\ell }$.
\end{corollary}

Since the locations can increase, it may be that, in the modified solution, the last items of $V^*$ do not fit in the knapsack.
To ensure that all items of $V^*$ are packed, we define an adjusted vector of interval lengths $c'$.
Let $W$ be the total length of intervals in the optimal solution, that is,~$W = \sum_{j = 1}^{|V^*| + 1} c^*_j$.
We consider each interval $j$ in order and set $c'_j = c_j$ until the interval $k$ at which the cumulative length is at least $W$. Then, we set the length of the next adjusted intervals to zero.
More precisely, let~$k$ be the first index such that~$\sum_{\ell = 1}^{k}c_\ell \ge W$. If no such index exists, we set~$c' = c$, otherwise, for each interval~$j$, define
\begin{equation}\label{eq:adjusted}
\textstyle
 c'_j =
\begin{cases}
 c_j    & \text{if } j < k, \\
 W - \sum_{\ell = 1}^{k-1}c_\ell, & \text{if } j = k, \\
0    & \text{if } j > k.
\end{cases}
\end{equation}
See Figure~\ref{fig:exemplo_vetor_c}.


Let $h'_j$ be the end location of interval~$j$ in the modified solution corresponding to the adjusted interval lengths~$c'$.
As before, the adjusted end locations cannot increase by much. The proofs of Lemma~\ref{lem:altura} and Corollary~\ref{cor:cumulative} are in Appendix~\ref{appendix:B}.

\usetikzlibrary {patterns.meta}
\pgfdeclarepattern{
 name=hatch,
 parameters={\hatchsize,\hatchangle,\hatchlinewidth},
 bottom left={\pgfpoint{-.1pt}{-.1pt}},
 top right={\pgfpoint{\hatchsize+.1pt}{\hatchsize+.1pt}},
 tile size={\pgfpoint{\hatchsize}{\hatchsize}},
 tile transformation={\pgftransformrotate{\hatchangle}},
 code={
  \pgfsetlinewidth{\hatchlinewidth}
  \pgfpathmoveto{\pgfpoint{-.1pt}{-.1pt}}
  \pgfpathlineto{\pgfpoint{\hatchsize+.1pt}{\hatchsize+.1pt}}
  \pgfpathmoveto{\pgfpoint{-.1pt}{\hatchsize+.1pt}}
  \pgfpathlineto{\pgfpoint{\hatchsize+.1pt}{-.1pt}}
  \pgfusepath{stroke}
 }
}

\tikzset{
 hatch size/.store in=\hatchsize,
 hatch angle/.store in=\hatchangle,
 hatch line width/.store in=\hatchlinewidth,
 hatch size=5pt,
 hatch angle=0pt,
 hatch line width=.5pt,
}

\definecolor{babyblue}{rgb}{0.54, 0.81, 0.94}
\definecolor{babyblueeyes}{rgb}{0.63, 0.79, 0.95}
\definecolor{bluegray}{rgb}{0.4, 0.6, 0.8}
\definecolor{mediumseagreen}{rgb}{0.24, 0.7, 0.44}

\begin{figure}[h]
 \centering
 \resizebox{.8\textwidth}{!}{%
  \begin{tikzpicture}

   \begin{scope}[start chain=going,node distance=10mm]
    \draw[black, fill=none] (-3, -9) rectangle ++(2, 8);
    \draw[draw=none, pattern=north east lines, pattern color=mediumseagreen] (-2.9, -1) rectangle ++(1.8, 0) node[pos=.5, xshift=-1.5cm] {\LARGE$c^*_1$};
    \draw[babyblue, pattern=hatch, pattern color=babyblue] (-2.9, -2.1) rectangle ++(1.8, 1);
    \draw[draw=none, pattern=north east lines, pattern color=mediumseagreen] (-2.9, -2.7) rectangle ++(1.8, .5) node[pos=.5, xshift=-1.5cm] {\LARGE$c^*_2$};
    \draw[babyblue, pattern=hatch, pattern color=babyblue] (-2.9, -3.8) rectangle ++(1.8, 1);
    \draw[draw=none, pattern=north east lines, pattern color=mediumseagreen] (-2.9, -3.8) rectangle ++(1.8, 0) node[pos=.5, xshift=-1.5cm] {\LARGE$c^*_3$};
    \draw[babyblue, pattern=hatch, pattern color=babyblue] (-2.9, -4.9) rectangle ++(1.8, 1);
    \draw[draw=none, pattern=north east lines, pattern color=mediumseagreen] (-2.9, -6) rectangle ++(1.8, 1) node[pos=.5, xshift=-1.5cm] {\LARGE$c^*_4$};
    \draw[babyblue, pattern=hatch, pattern color=babyblue] (-2.9, -6.9) rectangle ++(1.8, .8);
    \draw[draw=none, pattern=north east lines, pattern color=mediumseagreen] (-2.9, -8.2) rectangle ++(1.8, 1.2) node[pos=.5, xshift=-1.5cm] {\LARGE$c^*_5$};
    \draw[babyblue, pattern=hatch, pattern color=babyblue] (-2.9, -8.9) rectangle ++(1.8, .6);
    \draw[draw=none, pattern=north east lines, pattern color=mediumseagreen] (-2.9, -8.9) rectangle ++(1.8, 0) node[pos=.5, xshift=-1.5cm] {\LARGE$c^*_6$};

    \draw[black, fill=none] (3.5, -9) rectangle ++(2, 8);
    \draw[draw=none, pattern=north east lines, pattern color=mediumseagreen] (3.6, -1) rectangle ++(1.8, 0) node[pos=.5, xshift=-1.5cm] {\LARGE$c_1$};
    \draw[babyblue, pattern=hatch, pattern color=babyblue] (3.6, -2.1) rectangle ++(1.8, 1);
    \draw[draw=none, pattern=north east lines, pattern color=mediumseagreen] (3.6, -3.2) rectangle ++(1.8, 1) node[pos=.5, xshift=-1.5cm] {\LARGE$c_2$};
    \draw[babyblue, pattern=hatch, pattern color=babyblue] (3.6, -4.3) rectangle ++(1.8, 1);
    \draw[draw=none, pattern=north east lines, pattern color=mediumseagreen] (3.6, -4.3) rectangle ++(1.8, 0) node[pos=.5, xshift=-1.5cm] {\LARGE$c_3$};
    \draw[babyblue, pattern=hatch, pattern color=babyblue] (3.6, -5.4) rectangle ++(1.8, 1);
    \draw[draw=none, pattern=north east lines, pattern color=mediumseagreen] (3.6, -6.7) rectangle ++(1.8, 1.2) node[pos=.5, xshift=-1.5cm] {\LARGE$c_4$};
    \draw[babyblue, pattern=hatch, pattern color=babyblue] (3.6, -7.6) rectangle ++(1.8, .8);
    \draw[draw=none, pattern=north east lines, pattern color=mediumseagreen] (3.6, -8.2) rectangle ++(1.8, .5) node[pos=.5, xshift=-1.5cm, yshift=-.15cm] {\LARGE$c_5$};
    \draw (5.3,-8.3) node[right, xshift=.5cm] {\LARGE$W$} -- ++(0.4,0);
    \draw[fill=white, babyblue, pattern=hatch, pattern color=babyblue] (3.6, -9.2) rectangle ++(1.8, .6);
    \draw[draw=none, pattern=north east lines, pattern color=mediumseagreen] (3.6, -9.2) rectangle ++(1.8, 0) node[pos=.5, xshift=-1.5cm] {\LARGE$c_6$};

    \draw[black, fill=none] (10, -9) rectangle ++(2, 8);
    \draw[draw=none, pattern=north east lines, pattern color=mediumseagreen] (10.1, -1) rectangle ++(1.8, 0) node[right, pos=.5, xshift=1cm] {\LARGE$c'_1 = c_1$};
    \draw[babyblue, pattern=hatch, pattern color=babyblue] (10.1, -2.1) rectangle ++(1.8, 1);
    \draw[draw=none, pattern=north east lines, pattern color=mediumseagreen] (10.1, -3.2) rectangle ++(1.8, 1) node[right, pos=.5, xshift=1cm] {\LARGE$c'_2 = c_2$};
    \draw[babyblue, pattern=hatch, pattern color=babyblue] (10.1, -4.3) rectangle ++(1.8, 1);
    \draw[draw=none, pattern=north east lines, pattern color=mediumseagreen] (10.1, -4.3) rectangle ++(1.8, 0) node[right, pos=.5, xshift=1cm] {\LARGE$c'_3 = c_3$};
    \draw[babyblue, pattern=hatch, pattern color=babyblue] (10.1, -5.4) rectangle ++(1.8, 1);
    \draw[draw=none, pattern=north east lines, pattern color=mediumseagreen] (10.1, -6.7) rectangle ++(1.8, 1.2) node[right, pos=.5, xshift=1cm] {\LARGE$c'_4 = c_4$};
    \draw[babyblue, pattern=hatch, pattern color=babyblue] (10.1, -7.6) rectangle ++(1.8, .8);
    \draw[draw=none, pattern=north east lines, pattern color=mediumseagreen] (10.1, -8.2) rectangle ++(1.8, .5) node[right, pos=.5, xshift=1cm] {\LARGE$c'_5 = W - \sum_{\ell = 1}^{k-1}c_\ell$};
    \draw[babyblue, pattern=hatch, pattern color=babyblue] (10.1, -8.9) rectangle ++(1.8, .6);
    \draw[draw=none, pattern=north east lines, pattern color=mediumseagreen] (10.1, -9) rectangle ++(1.8, 0) node[right, pos=.5, xshift=1cm] {\LARGE$c'_6 = 0$};

    \draw (-2, -9.5) node[] {\LARGE$\opt$};


    \draw (11, -9.5) node[] {};

   \end{scope}

  \end{tikzpicture}
 }
 \caption[Examples of vectors of intervals.]{The first solution is an optimal packing~$\opt$ where, in blue (crossed diagonal lines) are the items of~$V^*$, in green (parallel diagonal lines) the area of intervals between them in $\opt$, and~$c^*$ corresponds to the vector of sizes of the intervals. The second solution is packing the same set of items in $\opt$ considering the intervals as a vector $c$ with multiples of $\e$. And the third solution is a packing of the same set of items using~$c'$, obtained by modifying~$c$ according to~\eqref{eq:adjusted}.\label{fig:exemplo_vetor_c}}
\end{figure}
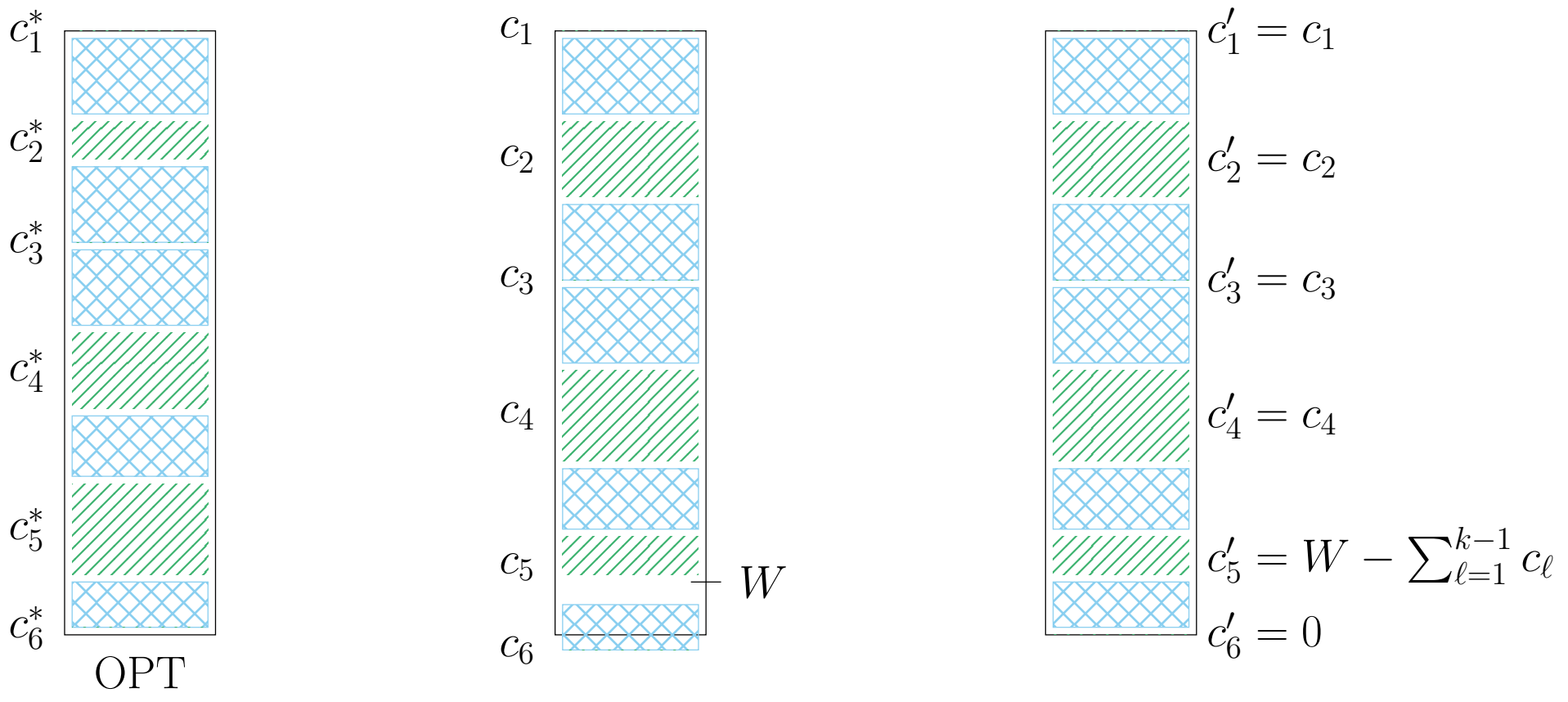

\begin{lemma}\label{lem:altura}
 For every interval~$j$, we have ${0 \le h'_j - h^*_j < \e}$.
\end{lemma}

\begin{corollary}\label{cor:cumulative}
 For every interval~$j$, we have
${ \sum_{\ell = 1}^j c'_\ell \ge \sum_{\ell = 1}^j c^*_\ell }$.
\end{corollary}

\bigskip

Now, for some subset of items $S$, define $f(S)$ as the total profit of items in $S$ when each item $i \in S$ is placed at location $h^*_i$, i.e., define ${f(S) = \sum_{i \in S} \gionly(h^*_i) v_i}$.
Analogously, define~${f'(S) = \sum_{i \in S} \gionly(h'_i) v_i}$ as the total profit of $S$ according to locations~$h'$.
Also, let $R^* = S^* \setminus V^*$. Thus, the value of the optimal solution is~${f(S^*) = f(V^*) + f(R^*)}$.

Because the start locations of items in $V^*$ change by at most $\e$ and the profit functions are relatively continuous, the loss in the solution value is bounded by a fraction of the optimal value.

\begin{lemma}\label{lem:val-v}
$f'(V^*) \ge (1-\delta) f(V^*)$.
\end{lemma}

\begin{proof}
 Consider some item $i \in V^*$. By Lemma~\ref{lem:altura}, we have $h'_i \le h^*_i + \e$.
 Since $\gionly$ is relatively continuous,
$
\gi{h'_i} \ge \gi{h^*_i + \e} \ge (1 - \delta)\gi{h^*_i} .
$
 Therefore,
\begin{align*}
 f'(V^*)
 = \sum_{i \in V^*} \gi{h'_i} v_i
\ge \sum_{i \in V^*} (1-\delta)\gi{h^*_i} v_i
 = (1-\delta) f(V^*).
\tag*{\qedhere}
\end{align*}
\end{proof}

Next, we tackle the remaining items.
Let $U$ be the set of items not in $V^*$ and whose \rev{weight} is smaller than~$\e^2$.
We consider the residual problem of allocating items from~$U$ in the intervals corresponding to~$c'$.
We want to find an allocation that approximates the profit for items in~$R^*$ in the optimal solution.
To do this, we will build and solve an instance of the \GAPs~(\GAP).

A \GAP instance consists of a set of bins, where each bin~$j$ has capacity~$C_j$, and a set of items, where each item~$i$ has value~$p_{ij}$ and \rev{weight}~$w_{ij}$ when added to bin~$j$.
The objective is to maximize the value of added items without exceeding any bin's capacity. \rev{When a constant bounds the number of bins, \GAP admits a PTAS~\citep{chekuri2005polynomial,frieze1984approximation}.\footnote{\citet{chekuri2005polynomial} state that ``a PTAS for this case exists even for instances of \GAP (implicit in earlier work)'', crediting the result to the $m$-dimensional knapsack approximation scheme of \citet{frieze1984approximation}. Neither paper gives a proof stated explicitly for \GAP under that name; the PTAS is only implicit in the construction of \citeauthor{frieze1984approximation}, as noted by \citeauthor{chekuri2005polynomial} themselves.}}

We build a \GAP instance as follows. For each nonempty interval $j$, create $q_j = \lceil c'_j/\e \rceil$ bins, which are denoted by $\{j^1, j^2, \dots, j^{q_j}\}$.
Each bin has capacity~$\e$, with possibly the exception of the last bin, whose capacity is
$
c'_j - \e (q_j - 1).
$
Observe that the set of bins corresponds to a partition of the intervals into sub-intervals.
Since the length of all but one interval is a multiple of $\e$, at most one bin has a capacity smaller than $\e$, corresponding to the last bin of the last nonempty interval; for this interval, the number of bins can exceed its length divided by~$\e$ by at most one, due to the ceiling.
Therefore, the number of bins is bounded by $1/\e + 1 \le 2/\e$, since $\e \le \delta < 1$.

For each $i \in U$, we consider a corresponding item of the $\GAP$ instance.
The weight and value of an item~$i$ when added to a bin $j^k$ are, respectively,
\begin{align}
 w_{ij^k} & = \s{i} \label{eq:weight}, \\
 p_{ij^k} & = \gi{h'_j -c'_j + k\e} v_i. \label{eq:value}
\end{align}
Note that the profit of an item~$i$ when added to bin~$j^k$ is the profit of the item when placed at the end location of the corresponding sub-interval. This concludes the construction of the instance.
The following lemma gives a lower bound on the optimal value of the $\GAP$ instance.

\begin{lemma}\label{lem:gaplower}
 An optimal solution for the $\GAP$ instance has value at least $(1-\delta)\cdot(f(R^*) - 2\e f(V^*))$.
\end{lemma}

\begin{proof}
 We build a feasible solution for the $\GAP$ instance by iteratively ``moving up'' items of~$R^*$ in the order that they appear in the optimal solution $(S^*, h^*)$.
 First, consider a list of bins ordered by their corresponding sub-intervals and make the first bin the current one.
 For each item $i \in R^*$, we place $i$ at the first free location of the sub-interval corresponding to the current bin if it fits, and, if not, we discard~$i$, close the bin, and make the next bin current.

 Consider a non-discarded item $i$. Let $j^k$ be the bin to which $i$ is assigned, and $j^*$ be the interval index containing $i$ in the optimal solution~$(S^*, h^*)$.
 Let $d$ be the total \rev{weight} of items of~$R^*$ that precede~$i$ in the optimal solution.
 Since~$i$ is contained in interval~$j^*$ and $\s{i} > 0$, we have $\sum_{\ell = 1}^{j^*} c^*_\ell > d$.
 Also, the \rev{weight} of these items (including the discarded ones) fully occupies the closed bins, thus $\sum_{\ell = 1}^{j-1} c'_\ell \le d$.
 Combining this inequality with Corollary~\ref{cor:cumulative}, we get
\[
\sum_{\ell = 1}^{j^*}
 c^*_\ell
 >
\sum_{\ell = 1}^{j-1} c'_\ell
\ge
\sum_{\ell = 1}^{j-1} c^*_\ell
\]
 But then $j^* > j-1$, and $j^* \ge j$.

 Let $h'_i$ be the start location of $i$ after it is moved. Again, the total \rev{weight} of items in $R^*$ that precede~$i$ is at least $h'_i - \sum_{\ell = 1}^{j-1} \s{\ell}$ and at most $h^*_i - \sum_{\ell = 1}^{j^*-1} \s{\ell}$, thus $h'_i - \sum_{\ell = 1}^{j-1} \s{\ell} \le h^*_i - \sum_{\ell = 1}^{j^*-1} \s{\ell}$, and
\[
 h'_i
\le
 h^*_i + \sum_{\ell = 1}^{j-1} \s{\ell} - \sum_{\ell = 1}^{j^*-1}\s{\ell}
\le
 h^*_i.
\]

\medskip

 Next, we bound the value of a non-discarded item of the \GAP solution. Recall that the value of an item assigned to a bin corresponds to the profit of that item when placed at the end of the sub-interval corresponding to a bin. Since a sub-interval has length at most $\e$, each non-discarded item~$i$ has a value of at least
$
\gi{h'_i + \e} v_i \ge (1-\delta ) \gi{h'_i} v_i \ge (1-\delta ) \gi{h^*_i} v_i.
$

 To account for discarded items, let~$I^*$ be the set of~$1/\e^2$ items of~$V^*$ with the largest profits in the optimal solution~$(S^*, h^*)$, and let~$\Gmin$ denote the minimum profit among items of~$I^*$.
 Because there are~$1/\e^2$ items in~$I^*$ with profit at least~$\Gmin$, we know that~$\Gmin/\e^2 \le f(I^*) \le f(V^*)$, and thus~${\Gmin/\e \le \e f(V^*)}$.
 Recall that there are at most $2/\e$ bins; thus, at most $2/\e$ items of $R^*$ are discarded. Because each item in $R^*$ has profit at most $\Gmin$, the total profit of non-discarded items is at least $f(R^*) - 2\Gmin/\e \ge f(R^*) - 2\e f(V^*)$.
 Since the value of the constructed solution is at least a fraction $(1-\delta)$ of the profit of non-discarded items, the lemma follows.
\end{proof}

For each combination of set~$V'$ with up to $2/\e^2$ items, permutation of $V'$ and vector~$c$, we define~$h'$ accordingly, construct a \GAP instance and find a $(1-\delta)$-approximation for the instance using some known PTAS for \GAP.
The solution for this instance induces a set of items $R'$ to add to the knapsack and a corresponding vector of locations~$h'$.
We make the union $S' = V' \cup R'$ and check whether $(S', h')$ is a feasible solution. The algorithm outputs the feasible solution that maximizes~$f'(S')$.
A pseudo-code summarizing the whole PTAS is given in~Algorithm~\ref{alg:ptas}.

\begin{algorithm}[h]
\begin{small}
\caption{PTAS for \PKP with a set of functions~$\calg$.\label{alg:ptas}}
\begin{algorithmic}[1]
\Procedure{PTAS-\PKP-$\calg_{\delta, \e}$}{$I, w, v, \gonly$}
\State Let $B = \{i \in I \mid \s{i} \geq \e^2 \}$.
\For{each $I' \subseteq I$ such that $|I'| \le 1/\e^2$}\label{line6:3}
\For{each $B' \subseteq B$ such that $|B'| \le 1/\e^2$}\label{line6:4}
\State Let $V' = I' \cup B'$.
\For{each permutation of $V'$}
\For{each vector~$c$ with $|V'|+1$ multiples of $\e$ s.t.
$\sum_{\ell = 1}^{|V'|+1}c_\ell \le 1$}
\State Let $W = 1 - \sum_{j \in V'} \s{j}$.
\State Compute the adjusted lengths $c'$ as in~\eqref{eq:adjusted}.
\State Set the location $h'_j$ of each $j \in V'$ according to $c'$.
\State For each $j = 1, 2, \dots, |V'|+1$, create bins $\{j^1, j^2, \dots, j^{q_j}\}$.
\State Let $U = I \setminus (V' \cup B)$.
\State For each $i \in U$ create an item with weight and value as in~\eqref{eq:weight} and~\eqref{eq:value}.
\State Create a \GAP instance with created items and bins.
\State Find a $(1-\delta)$-approximation of the \GAP instance.
\State Let $R'$ be the returned items with location $h'_i$ for each $i \in R'$.
\State Let $S' = V' \cup R'$ and compute the value of the solution $(S', h')$.
\EndFor
\EndFor
\EndFor
\EndFor
\State \Return a solution $(S', h')$ with maximum $f'(S')$.
\EndProcedure
\end{algorithmic}
\end{small}
\end{algorithm}

Lemma~\ref{lem:ptas_poly}, proved in Appendix~\ref{appendix:B}, bounds the running time of the algorithm, and Theorem~\ref{thm:ptas} combines it with the previous results.

\begin{lemma}\label{lem:ptas_poly}
 Algorithm~\ref{alg:ptas} runs in polynomial time.
\end{lemma}

\begin{theorem}\label{thm:ptas}
 Algorithm~\ref{alg:ptas} is a PTAS for \PKP with a set of functions~$\calg$.
\end{theorem}

\begin{proof}
 By Lemma~\ref{lem:ptas_poly}, the algorithm runs in polynomial time in the size of the instance.

 Since the algorithm tries each possible permutation of a set with up to $2/\e^2 $ elements, and each possible vector $c$ containing $|V^*| + 1$ multiples of $\e$, it considers the permutation of $V^*$ and the vector $c$ corresponding to an optimal solution $(S^*, h^*)$. Let $(S', h')$ be the solution constructed for these choices. Also, recall that the value of this solution is ${f'(S') = f'(V') + f'(R')}$, and the value of the optimal solution is ${f(S^*) = f(V^*) + f(R^*)}$.

 By Lemma~\ref{lem:val-v}, $f'(V') = f'(V^*) \ge (1-\delta) f(V^*)$.
 Also, the value $f'(R')$ is at least the value of the solution built for the \GAP instance.
 Since~$R'$ corresponds to a ${(1-\delta)}$-approximation for this instance, Lemma~\ref{lem:gaplower} implies that $f'(R') \ge (1-\delta) \cdot (1-\delta) (f(R^*) - 2\e f(V^*))$.
 Combining these inequalities, and recalling that $\e \le \delta$, we have:
\begin{align*}
 f'(S')
& \ge (1-\delta) f(V^*) + (1-\delta) \cdot (1-\delta) (f(R^*) - 2\e f(V^*)) \\
& \ge (1-\delta) f(V^*)  + (1-2\delta) f(R^*) - 2\delta f(V^*) \\
& = (1-3\delta) f(V^*) + (1-2\delta) f(R^*)     \\
& \ge (1-3\delta) (f(V^*) + f(R^*))     \\
& = (1-3\delta) f(S^*).
\end{align*}
 The second inequality follows by expanding the product: since~${f(R^*) \ge 0}$ and~${{(1-\delta)}^2 \ge 1 - 2\delta}$, the first resulting term is at least~${(1-2\delta)f(R^*)}$; and since~${f(V^*) \ge 0}$, ${{(1-\delta)}^2 \le 1}$ and~${\e \le \delta}$, the second one is at least~${-2\delta f(V^*)}$.
 Since the algorithm outputs the best feasible solution, this completes the proof of the theorem.
\end{proof}

\begin{acknowledgements}
A preliminary version of this work was presented as a brief announcement in LAGOS 2023~\citep{pedrosa2023positional}.
\rev{We thank the anonymous reviewers for their valuable comments that improved the presentation of this article. In particular, we thank the second reviewer for suggesting the reduction presented in Section~\ref{subsec:np-hardness-tardiness}.}
\end{acknowledgements}

\bibliographystyle{plainnat}      
\bibliography{main}

%
%
%





\clearpage
\appendix



%

\section{Omitted Proofs}\label{appendix:B}

\begin{proof*}[of Lemma~\ref{lem:statement0}]
\begin{footnotesize}
  \begin{flalign*}
 f(S)
  & = \sum\limits_{i = 1}^{r} \G{i}
 = \sum\limits_{i = 1}^{r}\s{i} \left(\L - \sum\limits_{j = 1}^{i - 1} \s{j}\right)                             \\
  & = \L\sum\limits_{i = 1}^{r}\s{i} - \frac{1}{2}\sum\limits_{i = 1}^{r}\sum\limits_{j = 1}^{r}\s{i} \s{j} + \frac{1}{2} \sum\limits_{i = 1}^{r}\s{i}^2            \\
  & = \L\sum\limits_{i = 1}^{r}\s{i} - \frac{1}{2}{\left(\sum\limits_{i = 1}^{r}\s{i}\right)}^2 + \frac{1}{2}\sum\limits_{i = 1}^{r}\s{i}^2               \\
  & = (mM + A)\sum\limits_{i = 1}^{r}\left(M + a_i\right) - \frac{1}{2}{\left(\sum\limits_{i = 1}^{r}\left(M + a_i\right)\right)}^2 + \frac{1}{2}\sum\limits_{i = 1}^{r}(M^2 + 2Ma_i + a_i^2) \\
  & = (mM + A)\left(rM + a(S)\right) - \frac{1}{2}{\left(rM + a(S)\right)}^2 + \frac{1}{2}(rM^2 + 2Ma(S) + a^2(S))                   \\
  & = \frac{1}{2}\left( r M \left( M - rM + 2mM + 2A\right) + 2 \left(\left(1 - r + m\right)M + A\right) a(S) - {a(S)}^2 + a^2(S)\right).
  \end{flalign*} 
\end{footnotesize}  
\end{proof*}

\begin{proof*}[of Lemma~\ref{lem:3}]
\rev{%
 First, suppose, for a contradiction, that~$|S| > m$ for some feasible solution~$S$. Then
$
 \sum_{i \in S} \s{i} > (m+1)M > mM + A = \L,
$
 so~$S$ is not feasible.

 Next, let~$S'$ be the set of~$m$ items with the smallest values~$a_i$. Then
\[
 \sum_{i \in S'}{\s{i}}
 = mM + \sum_{i \in S'}{a_i}
 \le mM + A
 = \L,
\]
 so~$S'$ is feasible.

 Now suppose that~$|S| \le m-2$ for some solution~$S$, and let~$\s{\max} = \max_{i \in {I \setminus S}}{\s{i}}$. Since~$a_i \le A$ for every~$i \in I$ (see the construction), we have~${\s{\max} = M + a_{\max} \le M + A}$, where~${a_{\max} = \max_{i \in I \setminus S}{a_i}}$. Then
\begin{flalign*}
 \sum_{i \in S} \s{i} & \leq (m-2)M + \sum_{i \in S}{a_i} \leq mM - 2M + 2A \\
     & = \L - 2M + A = \L - M - 8A^2 + A      \\
     & < \L - M - A \le \L - \s{\max},
\end{flalign*}
 so one more item can be added without exceeding the capacity, increasing the value of~$S$. Thus no solution with fewer than~$m-1$ items is optimal, and every optimal solution has size~$m-1$ or~$m$.

 It remains to show that any solution with~$m$ items has strictly larger value than any solution with~$m-1$ items. Let~$S'$ and~$S''$ be solutions with~$|S'| = m$ and~$|S''| = m-1$. Note that
\begin{footnotesize}
\begin{flalign*}
 f(S')
 & = \frac{1}{2}\left(mM (M - mM + 2mM + 2A) + 2 ((1 - m + m) M + A) a(S') - {a(S')}^2 + a^2(S')\right) \\
 & = \frac{1}{2}\left(mM (M + mM + 2A) + 2(M + A) a(S') - {a(S')}^2 + a^2(S')\right)     \\
 & > \frac{1}{2}\left(mM (M + mM + 2A)\right).
\end{flalign*}
\end{footnotesize}
 The inequality holds since~$2(M + A) > a(S')$.
 Also,
\begin{small}
\begin{flalign*}
 f(S'')
 & = \frac{1}{2}\Big( (m-1) M ( M - (m-1)M + 2mM + 2A)          \\
 & \phantom{=} + 2 ((1 - (m-1) + m)M + A) a(S'') - {a(S'')}^2 + a^2(S'')\Big)    \\
 & = \frac{1}{2}\left((m-1)M (2M + mM + 2A) + (4M + 2A)a(S'') - {a(S'')}^2 + a^2(S'')\right) \\
 & < \frac{1}{2}\left((m-1)M (2M + mM + 2A) + (4M + 2A)2A \right)       \\
 & = \frac{1}{2}\left((m-1)M (M + mM + 2A) +(m-1)M^2 + 8AM + 4A^2 \right)     \\
 & < \frac{1}{2}\left((m-1)M (M + mM + 2A) +(m+1)M^2 \right).
\end{flalign*}
\end{small}
 The first inequality holds because~${a(S'') < 2A}$ and ${a^2(S'') \le {a(S'')}^2}$, and the last inequality holds because~${8AM \le M^2}$ and ${4A^2 < M^2}$.
 Therefore,
\begin{flalign*}
 f(S') - f(S'')
 & > \frac{1}{2}\left( M (M + mM + 2A) - (m+1) M^2\right) \\
 & > \frac{1}{2}\left( (m + 1) M^2 - (m+1) M^2\right) = 0.
\end{flalign*}
 Hence every solution with~$m$ items beats every solution with~$m-1$ items, and every optimal solution has exactly~$m$ items.}\qed
\end{proof*}

\begin{proof*}[of Lemma~\ref{lem:altura0}]
 The proof is by induction on~$j$. The end location of the first interval increases by less than~$\e$; thus, the inequality holds for $j = 1$.
 Now, let~${j \ge 2}$, and suppose by induction that~$0 \le h_{j-1} - h^*_{j-1} < \e$.
 If~$c_j > 0$, interval $j$ has nonzero length, and~$c_j$ is the smallest nonnegative multiple of~$\e$ that is at least~$\gamma_j := h^*_j - (h_{j-1} + \s{j-1})$. Hence~$0 \le c_j - \gamma_j < \e$, that is,~$0 \le h_j - h^*_j < \e$.
 Otherwise, $c_j = 0$, so~$\gamma_j \le 0$ and $h_j = h_{j-1} + \s{j-1} \ge h^*_j$. Combined with~$h^*_j \ge h^*_{j-1} + \s{j-1}$,
\begin{align*}
 0 \le h_j - h^*_j
 \le (h_{j-1} + \s{j-1}) - (h^*_{j-1} + \s{j-1})
 = h_{j-1} - h^*_{j-1} < \e. \tag*{\qed}
\end{align*}
\end{proof*}

\begin{proof*}[of Corollary~\ref{cor:cumulative0}]
 By direct calculation and using Lemma~\ref{lem:altura0},
\begin{align*}
 \sum_{\ell = 1}^j c_\ell
 =
 h_j - \sum_{\ell = 1}^{j-1} \s{\ell}
 \ge
 h^*_j - \sum_{\ell = 1}^{j-1} \s{\ell}
 =
 \sum_{\ell = 1}^j c^*_\ell.\tag*{\qed}
\end{align*}
\end{proof*}

\begin{proof*}[of Lemma~\ref{lem:altura}]
\rev{%
 Recall that the end location of an interval~$j$ is the cumulative length of the intervals up to~$j$ plus the total weight of the items of~$V^*$ that precede it, that is,
\begin{align*}
 h'_j = \sum_{\ell = 1}^{j} c'_\ell + \sum_{\ell = 1}^{j-1} \s{\ell},
 \qquad
 h_j = \sum_{\ell = 1}^{j} c_\ell + \sum_{\ell = 1}^{j-1} \s{\ell},
 \qquad\text{and}\qquad
 h^*_j = \sum_{\ell = 1}^{j} c^*_\ell + \sum_{\ell = 1}^{j-1} \s{\ell}.
\end{align*}
 By~\eqref{eq:adjusted}, we have~$\sum_{\ell = 1}^{j} c'_\ell = \min\bigl\{\sum_{\ell = 1}^{j} c_\ell,\, W\bigr\}$ for every interval~$j$. Indeed, if~$j < k$, then~$c'_\ell = c_\ell$ for every~${\ell \le j}$ and, by the minimality of~$k$, we have~${\sum_{\ell = 1}^{j} c_\ell < W}$; and if~$j \ge k$, then~${\sum_{\ell = 1}^{j} c'_\ell = W}$ while~${\sum_{\ell = 1}^{j} c_\ell \ge \sum_{\ell = 1}^{k} c_\ell \ge W}$. Therefore,
\begin{align*}
 h'_j = \min\Bigl\{h_j,\; W + \sum\nolimits_{\ell = 1}^{j-1} \s{\ell}\Bigr\}.
\end{align*}

 Since~${h'_j \le h_j}$ and, by Lemma~\ref{lem:altura0}, ${h_j < h^*_j + \e}$, the end location of interval~$j$ increases by less than~$\e$, that is,
\begin{align*}
 h'_j - h^*_j \le h_j - h^*_j < \e.
\end{align*}
 It remains to show that $h'_j \ge h^*_j$. By Lemma~\ref{lem:altura0}, we have $h_j \ge h^*_j$. Moreover, as $W \ge \sum_{\ell = 1}^{j} c^*_\ell$, we also have
\begin{align*}
 W + \sum_{\ell = 1}^{j-1} \s{\ell}
 \ge
 \sum_{\ell = 1}^{j} c^*_\ell + \sum_{\ell = 1}^{j-1} \s{\ell}
 = h^*_j.
\end{align*}
 As $h'_j$ is the minimum of these two quantities, it follows that $h'_j \ge h^*_j$, and thus $0 \le h'_j - h^*_j < \e$.}\qed
\end{proof*}

\begin{proof*}[of Corollary~\ref{cor:cumulative}]
 By direct calculation and using Lemma~\ref{lem:altura},
\begin{align*}
 \sum_{\ell = 1}^j c'_\ell
 =
 h'_j - \sum_{\ell = 1}^{j-1} \s{\ell}
 \ge
 h^*_j - \sum_{\ell = 1}^{j-1} \s{\ell}
 =
 \sum_{\ell = 1}^j c^*_\ell.\tag*{\qed}
\end{align*}
\end{proof*}

\begin{proof*}[of Lemma~\ref{lem:ptas_poly}]
 Since $|I'|$, $|B'|$, and $|V'|$ are bounded by a constant, the number of choices and permutations of $V'$ is polynomial. Moreover, the number of distinct vectors containing at most $\Oh(1/\e^2)$ multiples of~$\e$ and adding up to one is a constant.
 For each fixed permutation of $V'$ and choice of $c$, the running time to execute each step is polynomial, and thus the algorithm is polynomial.\qed
\end{proof*}

\end{document}